\documentstyle[12pt,psfig]{article}

\newcommand\f{\frac}
\newcommand\be{\begin{equation} }
\newcommand\ee{\end{equation} }

\begin{document}

\title{`Strange Stars' - have they been discovered ?}
\author{A. R. Prasanna\dag \& Subharthi Ray\ddag\footnote{Present Address: Instituto de
Fisica, Universidade Federal Fluminense, Niteroi, RJ, Brasil; e-mail:sray@if.uff.br}}

\maketitle

\centerline{Physical Research Laboratory, Navarangpura, Ahmedabad 380 009, India}

\centerline{\dag prasana@prl.ernet.in}

\centerline{\ddag sray@prl.ernet.in}

\begin{abstract}

Recently there have been controversial claims about the nature of the
isolated compact star RXJ1856.5$-$3754, with one group claiming it to
be a strange star (Drake et al., 2002) while the other asserting it
to be a normal neutron star (Walter \& Lattimer, 2002). The
controversy arises mainly due to the distance estimate, which in turn
is used to resolve the measured angular diameter, and thus the
radiation radius $R_\infty$. In this we discuss the theoretical
constraints that appear from analysing the usual mass-radius relation
alongwith the redshift factors arising from the strong gravity
effects and possible lensing. Unless the distance estimate is
confirmed independently, without any uncertainty, it is premature to
come to any conclusion regarding the nature of this star.

\end{abstract}

Subject headings: star: RXJ1856.5$-$3754 -- strange star --
neutron star -- equation of state

\section{Introduction}
One of the hot topics in astrophysics of late has been the
possibility of the discovery of Strange Stars (stars composed of
u, d and s quark matter), through X-ray observations of the
candidate RX~J1856.35$-$3754 from the Chandra satellite by Drake
et al., (2002). Almost at the same time Walter \& Lattimer (2002)
have given a totally different picture of the same, claiming it
to be a normal neutron star. Though discussions of the
possibility of the existence of such stars dates back to the
eighties (Witten, 1984 and later by Alcock et al., 1986, Haensel
et al., 1986, etc.), with high resolution data from the X-ray
observations, recently observers have improved the accuracy for
ascertaining the dimensions of the compact objects associated
with the observed emissions (Walter et al., 1996 Walter \&
Mathews 1997 and Pons et al., 2001). During the last few years,
there have been several sources, which are considered neither as
black holes nor neutron stars, like the Her-X1, 4U~1820$-$30 (Dey
et al, 1998 and Bombaci, 1997), SAX~J1808.4$-$3658 (Li et al.,
1999a), 4U~1728$-$34 (Li et al., 1999b), PSR~0943$+$10 (Xu et
al., 1999). The absence of
spectral lines in the thermal components of the X-ray compact
sources from the observations by Chandra and the XMM-Newton,
led Xu (2002) to claim the existence of bare strange stars. While it is
extremely difficult to come to a conclusion regarding a candidate
being a black hole (for masses of the order of few solar masses),
it is slightly better for neutron stars, once proper estimates of
mass and radius are made along with the evidence for the
existence of a hard surface which could produce bursts.

\section{Determination of mass-radius relation}

The method of obtaining the crucial parameters from observational
data does depend upon the models adopted. One of the recent
candidates which has evoked a lot of interest in this context is
the source RXT1856.3-3754, an isolated compact star at a distance
of $\sim$ 140 pc in the outskirts of the RCrA dark molecular
cloud. Incidentally there are two conflicting claims regarding
this source with Drake et al. (2002) claiming it to be a `strange
star' with radiation radius $R_\infty$ ranging from 3.8-8.2 kms,
and mass $\sim$ 1.4 $M\odot$, whereas Walter \& Lattimer (2002)
claim it to be a neutron star with $R_\infty$ 15 km and $M \sim$
1.7 $M_\odot$. In this approach the radiation radius $R_\infty$ is
defined through the general relativistic relation
\begin{equation}
R_\infty = \frac{R}{\sqrt{1-\frac{2GM}{c^{2}R}}}
\label{eqn:rinfty}
\end{equation}
wherein $R$ and $M$ are the actual radius and mass of the star. It
is apparent that what is measured is $R_\infty$, through
observation of the angular diameter of the source, expressed as
$R_\infty/D$ where $D$ is the distance of the source. Leaving
aside the part of ambiguities and uncertainties in the measurement
of $D$, there seems to be very little consensus among observers
about the distance measurement in this case.

However, what needs to be carefully looked into, is the crucial
implications of the `measured' value for determining the mass and
radius of the star, which decides whether it is a neutron star or
an exotic strange star. As the relevant formulae used takes into
account the crucial redshift factor due to strong gravity effects,
it is also important to consider other possible general
relativistic effects for photon trajectories. In this context one
should take into account `the self-lensing effect due to the
sources' own gravitational field' of the emitted light rays close
to the central star. Nollert et al. (1989) have discussed this
effect while analyzing the relativistic `looks' of a neutron star,
who clearly points out the relevance of self-lensing while
analyzing the data from a compact source. Accordingly, they point
out that if $I_v$ is the specific intensity of the observed
radiation then $I_{vs}$ and $I_{v\infty}$ are related through the
equation
\begin{equation}
I_{v\infty} = I_{vs} \left( 1 - \f{2GM}{Rc^2} \right)^{3/2}
\label{eqn:intensity}
\end{equation} Hence, if one considers this relation for the specific
intensity and evaluate the relation between the radii $R_\infty$
and $R_s$ for a black body emission one finds the relation to be

\begin{equation}
R^2_\infty = R^2_s (1+Z) = R^2_s / \left( 1 -
\f{2GM}{R_sc^2}\right)^{1/2}
\label{eqn:redshift}
\end{equation} and not the one used
by Drake et al. or Walter et al. Before proceeding further with
the estimates one needs to consider few other aspects of general
relativity. It is well known that $R = 2MG/c^2$ denotes the event
horizon and further $R = 3GM/c^2$ corresponds to the  circular
photon orbit, as well as the radius at which the centrifugal force
reversal occurs for particles on circular geodesics (Abramowicz
and Prasanna, 1990); Heyl, (2000), using this feature had obtained
constraints on neutron star radii for the case of Type 1 X-ray
burst sources. Before considering the authenticity of the
constraint one ought to realize that the effect of centrifugal
reversal occurs only for purely circular geodesics, whereas the
presence of even a small radial velocity for the accreting
particles would change its orbit to non-circular motion in which
case the centrifugal reversal occurs only if the central star is
rotating (Prasanna, 2001). In view of this, the constraint $R_s >
3M$ may not be always effective, whereas $R_\infty > 3M$ is indeed
a must for any observation.

Considering now equation (\ref{eqn:redshift}) which can be written
as

\begin{equation}
 R^5_s - R^4_\infty R_s + 2M R^4_\infty = 0
\label{eqn:quintic}
\end{equation}
a quintic equation in $R$. Unlike in the case of a cubic,
there is no way of expressing a general condition between
$R_\infty$ and $M$ for the existence of real roots. However, one
can find numerically that if $R_\infty
>$ 5.4954$M \sim$ 3.75$GM/c^2$ then there are two positive real
roots, for $R_s$ one of which certainly lies outside $3GM/c^2$,
the photon circular orbit.

Table 1 yields the location of the highest real positive root
$R_s$ for given $m$ and $R_\infty$ chosen to be 9m. As these are
the only consistent numbers, satisfying the defining equation one
has to constrain M-R relation as given by this. For comparison we
have also given the photon radius $3GM/c^2$ for the corresponding
mass, and one clearly sees that the actual radius of the star is
greater than $3GM/c^2$ for the chosen $R_\infty$.

It may be seen from the table that for the mass $M \sim$ 1.7
$M_\odot$, and $R_\infty$ = 15.3 kms, the actual stellar radius
$R_s \approx$ 13.65, a value larger than that obtained by Walter
and Lattimer. Another important thing to notice here is that the
redshift factor `$Z$' when calculated for the entire range of
values of $M$ and $R_s$ as given in Table 1 yields a value
$\approx$ 0.256, which lies within the range as given by Pons et
al. On the other hand, the ranges of $M$ and $R$ as obtained,
after including the lensing factor takes the star away from the
estimates of Drake et al.

If lensing is not taken into account then the two radii $R$ and $R_\infty$
are related through eq. (1), which may be re-written
as

\begin{equation}
{\rm R^3~-~R_\infty^2R~+~2MR_\infty^2 ~=~ 0}
\label{eqn:cubic}
\end{equation}

As R represents the true radius of the star of mass M it is
imperative that R$_\infty$ has to be such that the equation will
have real roots. As the last term is always positive, one of the
roots is always negative. Considering the discriminant ${\rm
4R_\infty^4(M^2~-~R_\infty^2/27),}$ it is clear that the other
two roots, when real, are equal if ${\rm R_\infty ~=~3\sqrt{3}M}$
and real but not equal for ${\rm R_\infty~>~3\sqrt{3}M}.$

 Table \ref{tab:quintic}
gives the locations of the outer real root for R for a given M, and the
photon orbit R$_P$  and the Radiation radius ${\rm R_\infty}$
respectively, for this case.

\begin{table}[htbp]
\begin{center}
\caption{Solutions of the quintic equation (Eqn.
\ref{eqn:quintic}) in R for different values of masses,
 taking into account the effect of lensing. } \vskip
.5cm
\begin{tabular}{|c|c|c|c|c|}
\hline
        Mass &    Radius  of  &  Photon  &  Radiation \\
        &       the star (R$_s$) & Radius &  Radius (R$_\infty$)\\
        (M$_\odot$)& (kms) & (kms) & (kms) \\
\hline
       0.4   &  3.21224 &   1.764 &     3.6\\
       0.5   &  4.0153  &   2.205 &     4.5\\
       0.6   &  4.81836 &   2.646 &     5.4\\
       0.7   &  5.62142 &   3.087 &     6.3\\
       0.8   &  6.42448 &   3.528 &     7.2\\
       0.9   &  7.22754 &   3.969 &     8.1\\
       1.0   &  8.03059 &   4.41  &     9.0\\
       1.1   &  8.83365 &   4.851 &     9.9\\
       1.2   &  9.63671 &   5.292 &     10.8\\
       1.3   &  10.4398 &   5.733 &     11.7\\
       1.4   &  11.2428 &   6.174 &     12.6\\
       1.5   &  12.0459 &   6.615 &     13.5\\
       1.6   &  12.849  &   7.056 &     14.4\\
       1.7   &  13.652  &   7.497 &     15.3\\
       1.8   &  14.4551 &   7.938 &     16.2\\
       1.9   &  15.2581 &   8.379 &     17.1\\
       2.0   &  16.0612 &   8.82  &     18.0\\
\hline
\end{tabular}
\end{center}
\label{tab:quintic}
\end{table}

\begin{table}[htbp]
\begin{center}
\caption{Solutions of Eqn. (\ref{eqn:cubic}) in R for
different values of masses, without taking into
account the effect of lensing. } \vskip .5cm
\begin{tabular}{|c|c|c|c|c|}
\hline
        Mass &    Radius  of  &  Photon  &  Radiation \\
        &       the star (R$_s$) & Radius &  Radius (R$_\infty$)\\
        (M$_\odot$)& (kms) & (kms) & (kms) \\
\hline
0.4 & 2.70758 & 1.764 & 3.6 \\
0.5 & 3.38448 & 2.205 & 4.5 \\
0.6 & 4.06138 & 2.646 & 5.4 \\
0.7 & 4.73827 & 3.087 & 6.3 \\
0.8 & 5.41517 & 3.528 & 7.2 \\
0.9 & 6.09206 & 3.969 & 8.1 \\
1.0 & 6.76896 & 4.41  & 9.0 \\
1.1 & 7.44586 & 4.851 & 9.9 \\
1.2 & 8.12275 & 5.292 & 10.8\\
1.3 & 8.79965 & 5.733 & 11.7 \\
1.4 & 9.47654 & 6.174 & 12.6 \\
1.5 & 10.1534 & 6.615 & 13.5 \\
1.6 & 10.8303 & 7.056 & 14.4 \\
1.7 & 11.5072 & 7.497 & 15.3 \\
1.8 & 12.1841 & 7.938 & 16.2 \\
1.9 & 12.861  & 8.379 & 17.1 \\
2.0 & 13.5379 & 8.82  & 18.0\\
\hline
\end{tabular}
\end{center}
\label{tab:cubic}
\end{table}

\section{Neutron star and Strange star equation of states}

There are many neutron star equation of states (EOS) which give
mass-radius relation over a wide range when fed into the TOV
equation. Almost all of the EOSs are calculated by considering
either the relativistic Dirac-Brueckner-Hartree-Fock models, or
the relativistic field theoretical models, or the
non-relativistic potential models. Also, some models have been
considered with the possibility of the stellar core possessing a
Bose-Einstien condensate of the negative kaons (Kaplan \& Nelson,
1986 and Thorsson et al., 1994). We have chosen only four neutron
star EOSs just to compare them with the strange star models. Also
we saw from literature that there are a wide variety of neutron
star EOSs, but none of them come so close to the radius within 8
kms.

\begin{figure}[htbp]
\centerline{\psfig{file=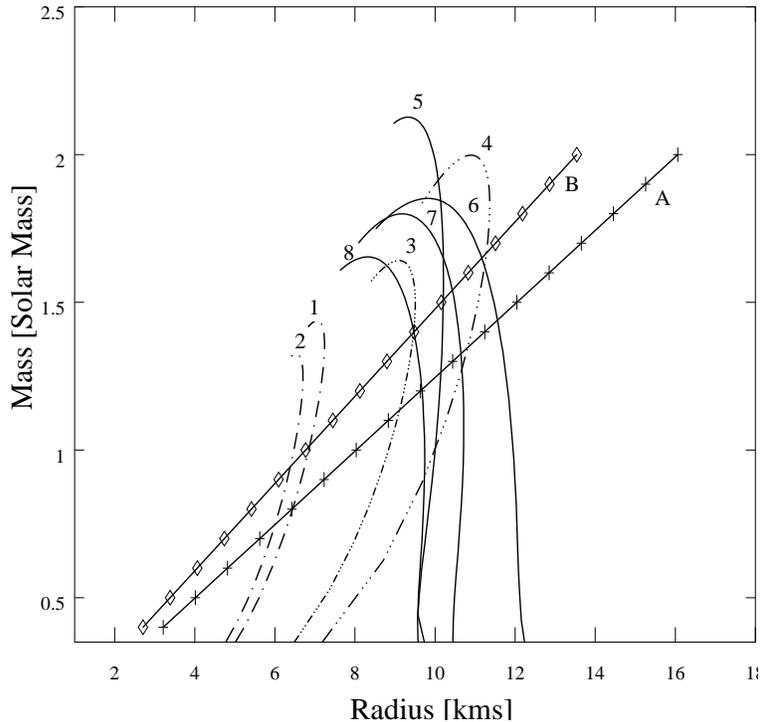,width=10cm}}
\caption{The lines labelled A \& B are the solutions of the roots of
Eqn.(\ref{eqn:quintic}) and Eqn.(\ref{eqn:cubic}) respectively.
It shows clearly that when the mass of the star is
low (say around 1 M$_\odot$), then the possibility of it being a neutron
star is completely ruled out and it is a strange star. However if the
mass is more than 1.5 M$_\odot$, then it is a neutron star. So, the
nature of the star cannot be established until an approximate value of
the mass of the star is provided.} \label{rxj:fig}
\end{figure}

The curve labelled 7 in Fig. (\ref{rxj:fig}) is due to Lorentz,
Ravenhall and Pethick (1993). They considered a microscopic
Hamiltonian obtained by fitting a Skyrme-like energy density
functional to the values of the employed microscopic two body
potential V14 and the three body force TNI. These has been
earlier used by Friedman and Pandharipande (1981) in hypernated
chain techniques considering a range of densities and
temperatures. Wiringa Ficks and Fabrocini (1988) calculated an
EOS (labelled 5 in Fig. (\ref{rxj:fig})) using a Hamiltonian where
a two-nucleon potential that fits nucleon-nucleon scattering data
and deuteron properties has been employed, and also possesses an
explicit three-nucleon interaction term. They calculated the
Hamiltonian using five combinations of the Argonne $v_{14}$,
Urbana $v_{14}$ two-nucleon potentials and the Urbana VII
three-nucleon potential. We have shown here only result with the
AV14+UVII. Curve labelled 6 is the EOS as per the Bethe Johnson
model (Bethe \& Johnson, 1974) who used realistic potentials in
their calculations, and curve labelled 8 is due to Pandharipande
(1971).

The idea of existence of exotic stars such as the strange quark
stars, has been a long term debate for astrophysicists and
particle physicists too. The laboratory scale lifetime for
deconfined u, d and s quarks is typically of the order of $\sim ~
10^{-24}$ sec, which is far away from the astrophysical scale of
stellar lifetime. So, a stable strange quark star model raised
questions about the existence of such stars. The scenario has
changed very much with the coming up of new generation x-ray
satellites. The analysis from observations now supply us new
constraints on the mass-radius relation of some of these compact
objects. With the advent of time, some of the new data for some sources and
their analysis (e.g., Haberl \& Titarchuk (1995), Bombaci (1997),
Dey et al. (1998), Li et al. (1999a \& 1999b), Xu et al. (1999), Kapoor
\& Shukre (2001), etc,) proved that the stars developed
from neutron star EOSs do not match with them, and they are more
compact. This leaves the only choice of considering them as the strange
quark stars.

    Ever since the strange star hypothesis was proposed by Witten
(1984), various models have been developed. Among all of them,
the most common one is that based on the MIT bag model. In this
phenomenological model the basic features of QCD, i.e., quark
confinement and asymptotic freedom, are built in. However in this
model, the deconfinement of quarks at high density is not
obvious. Preliminary calculations of strange stars using the bag
model has been done by Alcock et al. (1986) and Haensel et al.
(1986). A result for the bag model EOS for strange stars is shown
in the figure, where two cases for the  EOSs are given, one for
the $B = 94.9~ MeV/fm^3$ and strange quark masses of 0 MeV and
150 MeV for curves labelled 3 and 4 respectively.

Alternative to the MIT bag model, Dey et al., in 1998, derived an
EOS for strange matter which has asymptotic freedom built in and
describes deconfined quarks at high density and confinement at
zero density.  With a proper choice of the EOS parameters, this
model gives absolutely stable strange quark matter. This EOS was
used to calculate the structure of static strange stars and  the
mass-radius relations. Later, it was suggested (Li et al., 1999a)
that the millisecond X-ray pulsar SAX1808.4$-$3658 is a strange
star. This model which also explained the observed properties of
some other compact objects like the analysis of semi-empirical
mass-radius relations from the QPO observations in 4U~1728-34 (Li
et al., 1999b) as also the RXTE observations of Her~X-1 and
4U~1820-30 (Dey et al., 1998), leads to the suggestion that these
objects host strange stars. In Li et al., (1999a), two sets of
EOSs are used for two sets of parameters, namely SS1 and SS2.
The maximum gravitational masses are $M_{max}=1.437\,M_\odot$ for
SS1 (curve labelled 2) and $M_{max}=1.325\,M_\odot $ for SS2
(curve labelled 1). For different values of the mass parameter
$\nu$ in the D98 model, the stars can have a sequence of masses.
The calculations of Dey et al. (1998) has been done considering
zero temperature of the strange matter. Ray et al. (2000)
calculated the finite temperature effect on the quark stars
developed by Dey et al. (1998) and found that it sustains even
more mass for a particular radius of the star as compared to the
case of cold star.

\section{Discussions}

Fig.1 shows the lines of $M-R$ relation for the cases with lensing
(A) and without lensing (B) alongwith the curves depicting the
various equations of state both for neutron stars (solid lines) and
strange matter stars (dot dash lines). As is clear there can be
possibilities of identifying a star as of either category depending
upon the mass estimate and the corresponding radius, once the
measured radius $R_\infty$ is certain. Since the existence of the
real roots for $R$ does depend upon the relation between $R_\infty$
and $M$, it is very important that both these parameters are
estimated accurately and only then can one conclude about the nature
of the star, provided the associated point in Fig. 1 overlaps the
region covered by either of the types of equation of state. It is
also equally important to realise, whether the strong gravity effect
of lensing is to be taken into account or not. Either way, the
regions of overlap between the real roots and that of reasonable
equations of state are quite constrained and thus for a final
diagnosis of the star in question RXJ1856.35$-$3754, one needs to
have more unambiguous measurements of botyh $R_\infty$ and $M$. The
theoretical argument put forth above is purely of mathematical and
logistic in nature, which cannot be sidelined by any other
preferances.

\section{Acknowledgements}
We would like to thank Herman Marshall for a clarification through
correspondence,and John Miller for some useful comments. We would
also like to thank  B. G. A. Rao for some discussion regarding the
observational features.

\end{document}